\documentclass[onecolumn,nobibnotes,nofootinbib]{revtex4-1}
\usepackage{amsmath,amssymb,bm}
\usepackage{epsfig}
\usepackage{graphicx}
\usepackage{pstricks,pst-node,pst-text,pst-3d}
\usepackage{subfigure}

\newcommand{\vecr}{\bm{r}}

\newcommand{\vk}{\bm{k}}
\newcommand{\x}{\bm{x}}
\newcommand{\y}{\bm{y}}
\newcommand{\z}{\bm{z}}

\newcommand{\ab}{\bar\alpha}

\begin{document}
\title{\bf Momentum space saturation model for deep inelastic scattering and single inclusive hadron production}

 \author{E.\ A.\ F.\ Basso}
  \email{andre.basso@ufrgs.br}
 
 \author{M.\ B.\ Gay Ducati}
  \email{beatriz.gay@ufrgs.br}
  \affiliation{Instituto de F\'{\i}sica, Universidade Federal do Rio
                        Grande do Sul, Caixa Postal 15051, 91501-970 --- Porto Alegre, RS,
    Brazil}
    
 \author{E.\ G.\ de Oliveira}
  \email{emmanuel.de-oliveira@durham.ac.uk}
  \affiliation{Institute for Particle Physics Phenomenology, University of Durham, Durham, DH1 3LE, United Kingdom}

\begin{abstract}

We show how the AGBS model, originally developed for deep inelastic scattering applied to HERA data on the proton structure function, can also describe the RHIC data on single inclusive hadron yield for $d+Au$ and $p+p$ collisions through a new simultaneous fit. The single inclusive hadron production is modeled through the color glass condensate, which uses the quark(and gluon)--condensate amplitudes in momentum space. The AGBS model is also a momentum space model based on the asymptotic solutions of the BK equation, although a different definition of the Fourier transform is used. This aspect is overcome and a description entirely in transverse momentum of both processes arises for the first time. The small difference between the simultaneous fit and the one for HERA data alone suggests that the AGBS model describes very well both kind of processes and thus emerges as a good tool to investigate the inclusive hadron production data. We use this model for predictions at LHC energies, which agree very well with available experimental data.

\end{abstract}

\maketitle

\section{Introduction}

The understanding of the high energy behavior of the quantum chromodynamics (QCD) scattering amplitudes towards saturation has increased over the last decades. The pioneering work of Gribov, Levin, and Ryskin \cite{glr}, which aimed keep the high energy scattering amplitudes unitary in the small-$x$ limit of QCD by means of parton recombination, has started the field of investigation of the nonlinear evolution equations of the QCD amplitudes. The called GLR equation has also settled the presence into its solutions of a semi-hard energy dependent scale $Q_s(x)$ that determines the onset of the saturation effects in the evolved amplitudes, which posteriorly was called saturation scale. For probes with transversal momentum $k_t<Q_s$, the saturation effects are important to unitarize the scattering amplitudes. 

Since the GLR equation, other nonlinear evolution equations were developed to describe this high energy limit of QCD \cite{muelqiu,agl}, of which the simplest is the Balitsky--Kovchegov (BK) equation \cite{bal,kov1, kov2} for the dipole--target amplitude. BK equation is seen as a ``mean-field'' approximation once it neglects higher order correlations between the parton content of the amplitudes that are present in the so-called Balitsky \cite{bal} and JIMWLK \cite{jkmw97} equations, which consider the target as a Color Glass Condensate (CGC). 
The BK equation describes the rapidity evolution of a color dipole scattering by a target, being the dipole formed by a quark--antiquark pair with the transverse coordinates ${\x}$ and ${\y}$. 
In the fixed coupling case the scattering amplitude evolution in the two-dimensional position space reads
\begin{equation}\label{eq:bk}
\partial_Y {\cal N}_Y(\x,\y) = \ab \int d^2z\,\frac{|xy|^2}{|xz|^2|zy|^2}\left[{\cal N}_Y(\x,\z)
+{\cal N}_Y(\z,\y)-{\cal N}_Y(\x,\y) -{\cal N}_Y(\x,\z){\cal N}_Y(\z,\y) \right],
\end{equation}
where $|xy|^2=({\x} - {\y})^2$ is the dipole size, $Y=\ln 1/x$ the rapidity variable and $\ab = \alpha_sN_c/\pi$.

On the phenomenological side, the geometric scaling property \cite{gscaling} observed in deep inelastic scattering (DIS) data have determined the presence on the cross sections of the rapidity dependent saturation scale $Q_s$, as predicted from the GLR equation. This property establishes that the DIS cross sections depend on the Bjorken variable $x$ and on the virtuality $Q^2$ only through the combination $x/Q_s(x)$, where $Q_s$ is the saturation scale. An interesting feature between theory and phenomenology is the fact that only a saturation based model 
could describe the geometric scaling form of the DIS data at HERA \cite{gscaling}, 
giving us a strong empiric evidence of the saturation phenomena presence in the high energy collisions.

Even though numerical analyses of the fixed \cite{bk_num} and running coupling \cite{bkrc_num} versions of the BK equation have been performed, giving us insight about the solutions, the analytical solutions could not be found. This situation has changed with the discovery of the equivalence \cite{mp} between the BK equation and the Fisher--Kolmogorov--Petrovsky--Piscounov (FKPP) equation \cite{fkpp}. The later was largely studied in statistical physics \cite{saarloos, brunet-derrida} and is known to admit traveling waves as asymptotic solutions, which translates into the geometric scaling property of the QCD when applied to BK equation. The analysis is not dependent either on the form of the nonlinear term of the equation neither on the form of the linear kernel, so that the property of the traveling wave solutions is said to be universal. Furthermore, at asymptotic rapidities, when the wave reaches its traveling behavior, it loses the memory about the form of the initial conditions.

This procedure, called the traveling wave method of QCD, has received much theoretical effort, including studies of the higher order corrections of the BK kernel \cite{ps, bp, enberg}. However, there are few applications to phenomenology. The goal of this work is to extend a traveling wave based model \cite{agbs} to perform analyses both of the HERA DIS data on the proton structure function \cite{h1+zeus} and of the RHIC data on hadron yield \cite{brahms04,star06} through a simultaneous fit. 
In this way we can observe how the asymptotic solutions of the BK equation---as entering in the AGBS amplitude---behaves in nuclear collisions, with the parameters also limited to the DIS data at HERA.
The result of this analysis is used to describe the LHC $p_t$ distribution data for $p+p$ collisions \cite{cms-1,cms-2}.

The paper is organized as follows: in Section \ref{sec2} we review the traveling wave solutions of the BK equation and its modeling to DIS phenomenology. Section \ref{sec3} is devoted to the mathematical formulation and kinematics applied to DIS at HERA and to hadron production at RHIC. Section \ref{sec4} shows the data set and the results of the simultaneous fit as well as its predictions for LHC and Section \ref{sec5} presents our conclusions.

\section{The traveling wave solutions of BK equation and the AGBS model}\label{sec2}

For the BK equation in the one-dimensional form, when the impact parameter is neglected, the amplitude in (\ref{eq:bk}) takes the form ${\cal N}_Y(\x,\y)\equiv{\cal N}_Y(r)$, with $r=|xy|$. Thus it could be Fourier transformed through
\begin{equation}\label{eq:fourier}
 N(k,Y)=\frac{1}{2\pi} \int \frac{d^2r}{r^2}\,e^{i\mathbf{k}\cdot\mathbf{r}}\,{\cal N}(r,Y) = \int_0^\infty\frac{dr}{r}J_0(kr){\cal N}(r,Y),
\end{equation}
and the BK in momentum space reads \cite{kov1}
\begin{equation}\label{eq:bk_mom}
\partial_Y N(k,Y)=\bar{\alpha}\chi(-\partial_L) N(k,Y)-\bar{\alpha} N(k,Y)^{2},
\end{equation}
where
\begin{equation}\label{eq:kernel}
\chi(\gamma)=2\psi(1)-\psi(\gamma)-\psi(1-\gamma)
\end{equation}
is the BFKL \cite{bfkl} kernel and $L=\log(k^2/k_0^2)$, with a fixed soft scale $k_0$. 

In the diffusive approximation for this kernel, \textit{i.e.}, taking the Taylor expansion of the kernel up to second order, it can be shown \cite{mp} that the BK equation (\ref{eq:bk_mom}) belongs to the same universality class as the FKPP equation, which means that the BK equation admits traveling waves $N(L-v_g\bar{\alpha}Y)$ as solutions. The unique condition for the existence of these solutions is that the initial conditions $N(L,Y_0)$ decrease faster than $\exp(-\gamma_0 Y)$ for large $L$, with $\gamma_0>\gamma_c$. 
The color transparency property of QCD means that the scattering is weak when the dipole size is small, so that ${\cal N} \propto r^{2} $. Using (\ref{eq:fourier}) to Fourier transform this equation it gives $N \propto k^{-2\gamma_0}$, whit $\gamma_0 = 1$ at asymptotic $L$. The critical anomalous dimension $\gamma_c$ defines the group velocity $v_g(\gamma_c)$ of the asymptotic traveling wave. For the leading order BFKL kernel, $\gamma_c = 0.6275$ so that the condition $\gamma_0>\gamma_c$ is fulfilled.
 
Through the traveling wave method it is obtained the asymptotic part of the BK amplitude, valid for large $k$. For the fixed coupling case it reads \cite{mp}

\begin{equation}\label{eq:Ttail}
N\left(k,Y\right) \stackrel{k\gg Q_s}{\approx}
  \left(\frac{k^2}{Q_s^2(Y)}\right)^{-\gamma_c}\log\left(\frac{k^2}{Q_s^2(Y)}\right)
\exp\left[-\frac{\log^2\left(k^2/Q_s^2(Y)\right)}{2\bar{\alpha}\chi''(\gamma_c)Y}\right],
\end{equation}
where
\begin{equation}\label{eq:satscal}
Q_s^2(Y) = Q_0^2\exp\left( \lambda Y - \frac{3}{2\gamma_c}\log Y \right),
\end{equation}
is the saturation scale and $\lambda = \ab v_g = \ab \chi(\gamma_c)/\gamma_c$.

To make a phenomenological use of this asymptotic behavior, we need to add the infrared behavior for the amplitude in the region $k\ll Q_s$. This was done in \cite{agbs} using the Fourier transform of a two-dimensional step function, in order to unitarize the scattering amplitude in this saturated limit. The result is
\begin{equation}\label{eq:Tsat}
N(k,Y)\left(k\right) \stackrel{k\ll Q_s}{=} c-\log\left(\frac{k}{Q_s(Y)}\right).
\end{equation}
The equations (\ref{eq:Ttail}) and (\ref{eq:Tsat}) were analytically interpolated to form a momentum space parameterization of the dipole-target scattering amplitude, which reads

\begin{equation}\label{eq:agbs}
N(k,Y) = \left[ \log\left(\frac{k}{Q_s} + \frac{Q_s}{k}\right) + 1\right](1-e^{-T_\textrm{dil}}),
\end{equation}
where
\begin{equation}\label{eq:agbs_dil}
T_\textrm{dil} = \exp\left[ -\gamma_c\log\left(\frac{k^2}{Q_s^2(Y)}\right) - \frac{L_\textrm{red}^2 - \log^2(2)}{2\ab\chi^{''}(\gamma_c)Y}\right],
\end{equation}
with
\begin{equation}\label{eq:agbs_Qsat}
L_\textrm{red} = \log\left( 1 + \frac{k^2}{Q_s^2(Y)} \right)\qquad \textrm{and} \qquad Q_s^2(Y) = k_0^2\,e^{\lambda Y}.
\end{equation}
The equations above (\ref{eq:agbs}--\ref{eq:agbs_Qsat}) describe the AGBS model \cite{agbs} for the dipole--proton scattering amplitude, originally fitted to the HERA H1 and ZEUS collaborations noncombined data \cite{h1, zeus}. The model was also used to investigate the effects of the fluctuation in the gluon number during the dipole evolution \cite{IMM04,IT04,IM032,MS04} at the HERA energies \cite{agbs_fluct}.

\section{Phenomenology with AGBS model}\label{sec3}

\subsection{DIS at HERA}

A convenient frame to study the small-$x$ electron-proton deep inelastic scattering is the dipole frame, where the virtual photon emitted by the electron can split into a quark--antiquark pair that probes the target. In this way the photon--proton cross section factorizes in the photon wave function describing the probability of photon splitting into the $q\bar{q}$ pair and the dipole cross section describing the dipole interaction with the target \cite{nk-91, ahm94}:

\begin{equation}\label{eq:sig-fact}
\sigma_{T,L}^{\gamma^*p}(Q^2,Y) = \int d^2 r \int_0^1 dz |\Psi_{T,L}(r,z;Q^2)|^2 \sigma_\textrm{dip}(r,Y),
\end{equation} 
where $z$ is the momentum fraction carried by the quark. Neglecting the impact parameter of this interaction, \textit{i.e.}, taking the target as a homogeneous disk with fixed radius, the dipole cross section turns out to be a function of the dipole--target forward amplitude. In the case of the $e+p$ collisions we have
\begin{equation}
\sigma_\textrm{dip}(k,Y) = 2\pi R_p^2 N(k,Y),
\end{equation} 
where $R_p$ is the radius of the target proton.

The AGBS model was originally fitted to the proton structure function $F_2^p \propto \sigma_{T,L}^{\gamma^*p}$ at HERA, which was written from (\ref{eq:sig-fact}) in the momentum space as \cite{agbs}
\begin{equation}\label{eq:f2p}
F_2^p = \frac{Q^2R_p^2 N_c}{4\pi^2}\int_0^\infty \frac{dk}{k} \int_0^1 dz |\tilde{\Psi}_{T,L}(k^2,z;Q^2)|^2 N(k,Y),
\end{equation}
where $N(k,Y)$ refers to the AGBS dipole amplitude and $\tilde{\Psi}_{T,L} = \tilde{\Psi}_{T} + \tilde{\Psi}_{L}$ to 
the photon wave function, written in momentum space through
\begin{equation}\label{eq:psi_kspace}
\tilde{\Psi}_{T,L}(k,z) = \int \frac{d^2 r}{(2\pi)^2}\,e^{\imath\,\vk\cdot\vecr}r^2\Psi_{T,L}(r,z) = \int_0^\infty \frac{dr}{2\pi} r^3 J_0(kr)\Psi_{T,L}(r,z).
\end{equation}
Thus, we see from the convolution of the transforms of the photon wave function (\ref{eq:psi_kspace}) and of the dipole amplitude (\ref{eq:fourier}) that the whole transform of the proton structure function has the form of a two-dimensional Fourier (or Hankel) transform. As we shall see, this fact will be important in our goal to perform a simultaneous fit of the AGBS amplitude to the HERA and RHIC data.

\subsection{Inclusive hadron production at RHIC}

Within the frame of the Color Glass Condensate (CGC), it was proven that the differential cross section for single--inclusive forward hadron production in high energy 
collisions depends only upon dipoles through \cite{DHJ}
\begin{equation}\label{eq:had_yield}
\begin{split}
\frac{dN}{dy_h\,d^2p_t} = \frac{K}{(2\pi)^2}\int_{x_F}^1 \frac{dz}{z}&\left[x_1\,f_{q/p}(x_1,p_t^2)\tilde{N}_F\left(\frac{p_t}{z},x_2\right)D_{h/q}\left( z, p_t^2\right)\right.\\
& \left. + x_1\,f_{g/p}(x_1,p_t^2)\tilde{N}_A\left(\frac{p_t}{z},x_2\right)D_{h/g}\left( z, p_t^2\right) \right]\,,
\end{split}
\end{equation}
where a summation over quark flavors $q$ is understood. In this equation, $p_t$ and $y_h$ are the transverse momentum and rapidity of the produced hadron while $f_{i/p}$ and $D_{h/i}$ refer to the parton distribution function of the incoming nucleon and to the hadron fragmentation function respectively, which are considered at the scale $Q^2=p_t^2>1$ GeV${}^2$. Here we will use the CTEQ6 LO p.d.f's \cite{cteq6} and the LO KKP fragmentation functions \cite{kkp}. 
For light hadrons the finite mass effects can be neglected, so that the pseudorapidity $\eta$ and rapidity $y_h$ of the produced hadrons are similar $\eta \approx y_h$, giving the following kinematics: $x_F=\sqrt{m_h^2 + p_t^2}\exp(\eta_h)/\sqrt{S_{NN}}\approx p_t\exp(y_h)/\sqrt{S_{NN}}$, $x_2 = x_1\exp(-2y_h)$ and $x_1 = x_F/z$. 

The amplitude in momentum space is obtained through the Fourier transform of the dipole--proton forward scattering amplitude in the coordinate space
\begin{equation}\label{eq:ord_hnkl}
\tilde{N}_{F(A)}(k,Y) = \int d^2 \vecr\,e^{-\imath\vk\cdot\vecr}\,{\cal N}_{F(A)}(r,Y) = 2\pi\int dr\,r\, J_0(kr){\cal N}_{F(A)}(r,Y),
\end{equation}
where $N_F$ and $N_A$ are amplitudes in the fundamental and adjoint representations describing, respectively, the scattering of the quark and gluon content inside the projectile wavefunction by the target. The $N_F$ amplitude for quarks is obtained from the adjoint $N_A$ by the replacement $Q_s^2 \rightarrow (C_F/C_A)Q_s^2$, with $C_F/C_A = 4/9$.

In order to use the AGBS model in Eq.\ (\ref{eq:had_yield}) and fit it to RHIC data, we have to take into account the proper Fourier space where this equation was derived, given by Eq.\ (\ref{eq:ord_hnkl}). The AGBS amplitude $N(k)$, calculated with a different definition of Fourier transform given by Eq.\ (\ref{eq:fourier}), can be translated to: 
\begin{equation}\label{eq:agbs_nfs}
\tilde{N}(k) = 2\pi \left[ -\frac{d^2 N(k)}{dk^2} - \frac{1}{k}\frac{d N(k)}{dk} \right].
\end{equation}
So, performing the first derivatives of the AGBS amplitude (\ref{eq:agbs}) we get analytically an amplitude in the appropriate Fourier space to describe the RHIC data on single inclusive hadron production through (\ref{eq:had_yield}). 

Some conceptual comments about the fundamental and adjoint representations of the scattering amplitudes are in order. As pointed out in \cite{buw}, the dipole amplitude that enters in the calculation of DIS observables is an amplitude for quarks and so in the fundamental representation. Thus, in the dipole models applied to DIS, like the AGBS, the amplitude is a $N_F$ one without the $C_F/C_A$ factor mentioned before. To avoid this disagreement between the DIS and heavy ion amplitudes in our simultaneous fit to  HERA and RHIC data we use Eq.\ (\ref{eq:agbs_nfs}) as a model for $N_F$ and scale $Q_s^2$ $\rightarrow$ $C_A/C_F\,Q_s^2$ to obtain $N_A$.

\section{Results}\label{sec4}

\subsection{HERA and RHIC data simultaneous fit}

Both HERA proton structure function $F_2^p$ (\ref{eq:f2p}) (with the amplitude (\ref{eq:agbs})) and RHIC hadron yield (\ref{eq:had_yield}) (using the amplitude (\ref{eq:agbs_nfs})) data were used to simultaneously fit the AGBS model to DIS and heavy ion processes.

In this analysis, all the last combined HERA data measurements of the proton structure
function from H1 and ZEUS Collaborations \cite{h1+zeus} are fitted, within the
following kinematic range:
\begin{equation}
x\leq 0.01,
\end{equation}
\begin{equation}
0.1\leq Q^2 \leq 150\,\rm{ GeV}^2,
\end{equation}
which corresponds to 244 data points. Both ranges include values of $x$ low enough for the
analysis to be in the high energy regime, and values of $Q^2$ which allow us not to include
DGLAP corrections, although higher values of virtuality do not change significantly the results.

The RHIC data on single inclusive hadron production from BRAHMS \cite{brahms04} and STAR \cite{star06} collaborations were also fitted. They were considered in the $p_t$ range higher than $1$ GeV to guaranty that perturbative theory can be applied. To avoid contribution of large-$x$ in the target, we have focused our study in the forward ($y_h\geq 2$) rapidity region, although we also made the fit to the and mid-rapidity ($y_h=1$) region.
This gives us a total of 22 data points in the forward region analysis and 38 data points when the mid-rapidity region was included. In order to get the equal contributions from the HERA data and from the RHIC data, we have assigned a weight of $11$ ($6$ in the case including the region $y_h=1$) to the latter.

After all, for the simultaneous fit we have used 266 (only RHIC forward region) and 283 (RHIC forward and mid-rapidity regions) data points. Concerning the parameters, we keep fixed $\bar{\alpha}=0.2$ and $\gamma_c=0.6275$, whose value corresponds to the one obtained from the LO BFKL kernel. The other parameters in the amplitude---$\lambda$, $k_0^2$ and $\chi^{\prime\prime}(\gamma_c)$---are left to be free, as well as the proton radius $R_p$, which fixes the normalization of the dipole--proton cross section with respect to the dipole--proton amplitude for DIS at HERA, and the rapidity--dependent $K$ factors. We have to fix one of the $K$ factor to set the overall normalization; therefore, we set the value $K(y_h=4) = 0.7$, which was obtained from two different dipole models applied to the RHIC data \cite{dhj2,buw}. Only light quarks are considered and the values used for their masses were $m_{u,d,s}=140$ MeV.

\begin{table}[h!]
\footnotesize 
\begin{tabular}{|c@{\quad}|c@{\quad}|c@{\quad}|c@{\quad}|c@{\quad}|}
\hline

 $\chi^2/\mbox{d.o.f.}$ & $k_0^2$ ($\times 10^{-3}$) &  $\lambda$ &  $\chi^{\prime\prime}(\gamma_c)$  & $R$(GeV${}^{-1}$)  \\ [0.5ex] \hline \hline

0.903 & $ 1.13 \pm 0.024$ & $0.165 \pm 0.002$ & $7.488 \pm 0.081$ & $5.490 \pm 0.039$ \\ \hline

\end{tabular}
\caption{Parameters extracted from the fit to H1 and ZEUS combined data \cite{h1+zeus} on the proton structure function $F_2$ at HERA.}
\label{tab:DIS-res}
\end{table}

\begin{figure}[t!]
\centering
\subfigure[$d+Au$ collisions at RHIC for $y_h \geq 1.0$]{
\includegraphics[scale=0.33,angle=-90]{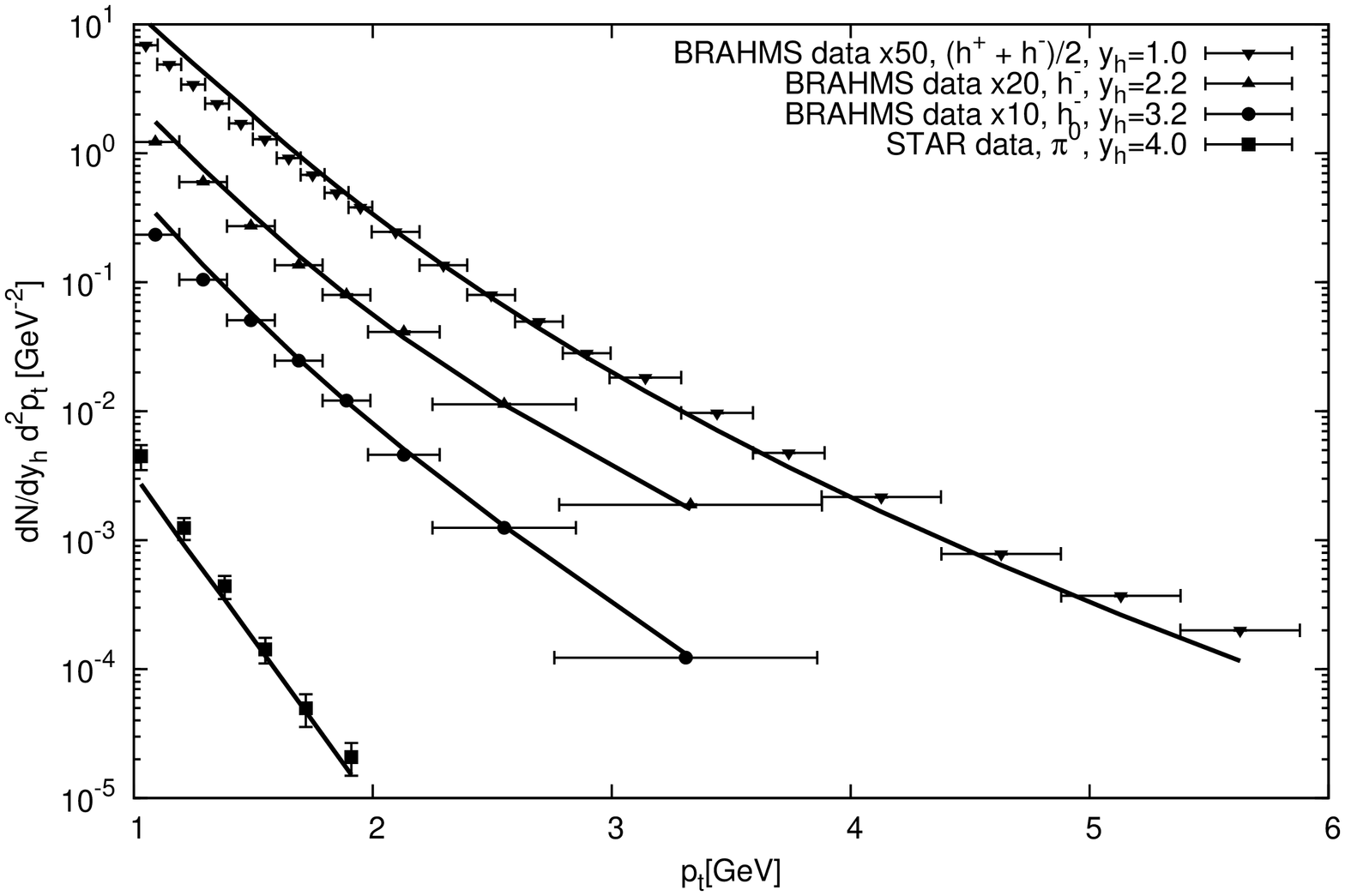}
\label{fig:agbs-dAu-yh1}
}
\subfigure[$d+Au$ collisions at RHIC for $y_h \geq 2.2$]{
\includegraphics[scale=0.33,angle=-90]{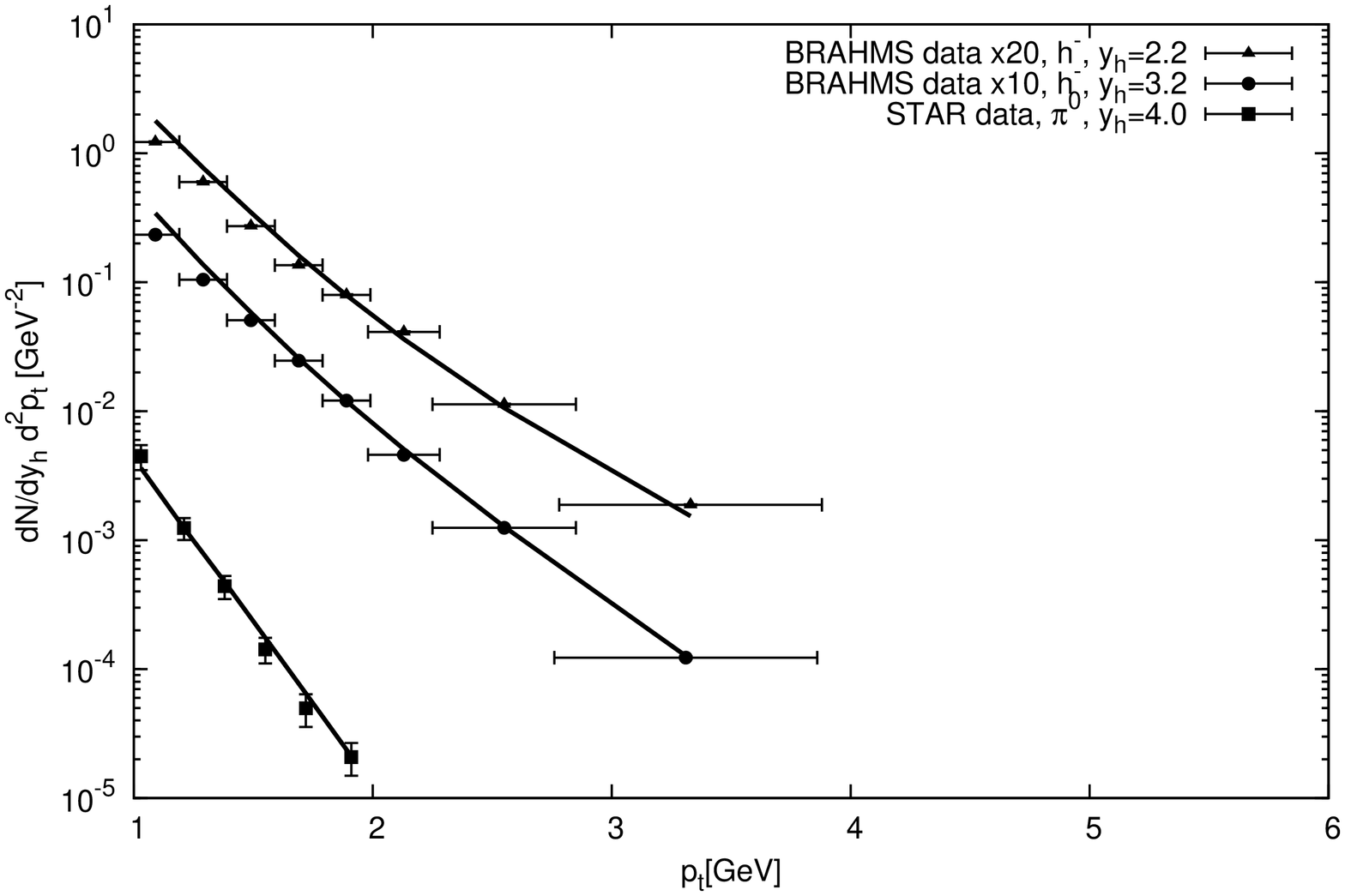}
\label{fig:agbs-dAu-yh2}
}
\label{fig:agbs_rhic_dAu}
\caption[]{Results for the RHIC charged hadron and $\pi^0$ yield for $d+Au$ collisions from the simultaneous fit of AGBS to RHIC \cite{brahms04,star06} and HERA  \cite{h1+zeus} data.}
\end{figure}

\begin{table}[h!]
\scriptsize 
\begin{tabular}{|c||c|c@{\quad}|c@{\quad}|c@{\quad}|c@{\quad}|c@{\quad}|c|c|c|c|}
\hline
 & $\chi^2/\mbox{d.o.f.}$ & $k_0^2$ ($\times 10^{-3}$) & $\lambda$ &  $\chi^{\prime\prime}(\gamma_c)$  & $R$(GeV${}^{-1}$) &  $K(y_h=1.0)$ & $K(y_h=2.2)$ & $K(y_h=3.2)$ & $K(y_h=4.0)$  \\ [0.5ex] \hline \hline
$y_h \geq 2.2 $ & 0.799 & $2.760 \pm 0.130$ & $0.190 \pm 0.003$ & \;\,$5.285 \pm 0.123$  & $4.174 \pm 0.053$  & -- & $2.816 \pm 0.110$ & $2.390 \pm 0.098$ & $0.7$ \\ \hline

$y_h \geq 1.0 $ & 1.056 & $1.660 \pm 0.137$ & $0.186 \pm 0.003$ & $6.698 \pm 0.223$ & $4.695 \pm 0.112$  & $6.172 \pm 0.379$ & $3.783 \pm 0.259$ & $3.256 \pm 0.226$ & $0.7$ \\ \hline

\end{tabular}
\caption{Parameters extracted from the simultaneous fit to HERA $F_2$ (H1 and ZEUS combined data \cite{h1+zeus}) and to the RHIC hadron yield for the $d+Au$ collisions (BRAHMS and STAR data \cite{brahms04,star06}). The \textit{first line} corresponds to the fit with the forward rapidity region of the RHIC data included, while in the \textit{second line} the mid-rapidity region of the RHIC data set was also considered.}
\label{tab:res}
\end{table}

First of all we performed a fit to DIS data, in order to verify the behavior of the AGBS model to the new combined H1 and ZEUS data \cite{h1+zeus}. The values of the parameters, shown in the table \ref{tab:DIS-res}, are close to those obtained in the original AGBS fit to the HERA data, although the recent data set is a little bit different and does not need the inclusion of a normalization uncertainty of 5\% in the H1 data. This fit will also serve as a guideline for our simultaneous fit, once DIS processes are free from model dependent PDF's and fragmentation functions that can interfere on the best values for the model parameters.

The simultaneous fit to DIS and hadron production in $d+Au$ collisions has shown a good agreement with the data set, mainly when only forward RHIC data was considered, as seen in the $\chi^2$ values of the table \ref{tab:res}. In the Figs.\ \ref{fig:agbs-dAu-yh1} and \ref{fig:agbs-dAu-yh2}  we see the description of the RHIC hadron yield for both lines of table \ref{tab:res}. The poorer description of the mid-rapidity region occurs because the target have not reached its gluon condensate state, when neither our amplitude nor the CGC formulation entering the Eq. (\ref{eq:had_yield}) are valid. The parameters values---even those got when the mid-rapidity data was included---shows that the AGBS model describes equally well the HERA and RHIC data, and the last improves the AGBS model. The saturation exponent $\lambda$ is slightly bigger than that coming from the HERA fit only, close to the expected value $\lambda = 0.2\div0.3$ demanded from NLO DIS \cite{triant_RG_lambda}. The saturation scale is just a little bit greater than that extracted from the DIS fit only, which have given us an unexpected small value for the nuclear saturation scale. However, $Q_s^2$ includes an additional factor $A_\textrm{eff}^{1/3}$ to account for the dipole interaction with the whole nuclear target and must be viewed as an average value. The value in the center of the nucleus should be higher and an analysis of the impact parameter influence on the AGBS amplitude will be important for the LHC energies, as seen from other saturation models \cite{tv}. We have used $A_\textrm{eff} = 18.5$ for $d+Au$ collisions.

Concerning the $K$ factors, we had to fix the pion one to normalize such factors as explained before. We used the value $K=0.7$ given by other two LO models \cite{dhj2, buw}. In fact, the first line of table \ref{tab:res} shows almost the same values of these LO models, with a factor of two for the hadron production ($y_h = 2.2$ and $y_h = 3.2$) as expected from LO models. We must note, however, that these factors are just normalization factors over the equation (\ref{eq:had_yield}) in the fitting procedure, describing the model uncertainties both on the DIS and hadron yield data, that actualy has 15\% of normalization uncertainty.

\begin{figure}[ht]
\centering
\subfigure[$p+p$ collisions at RHIC for $y_h \geq 1.0$]{
\includegraphics[scale=0.33,angle=-90]{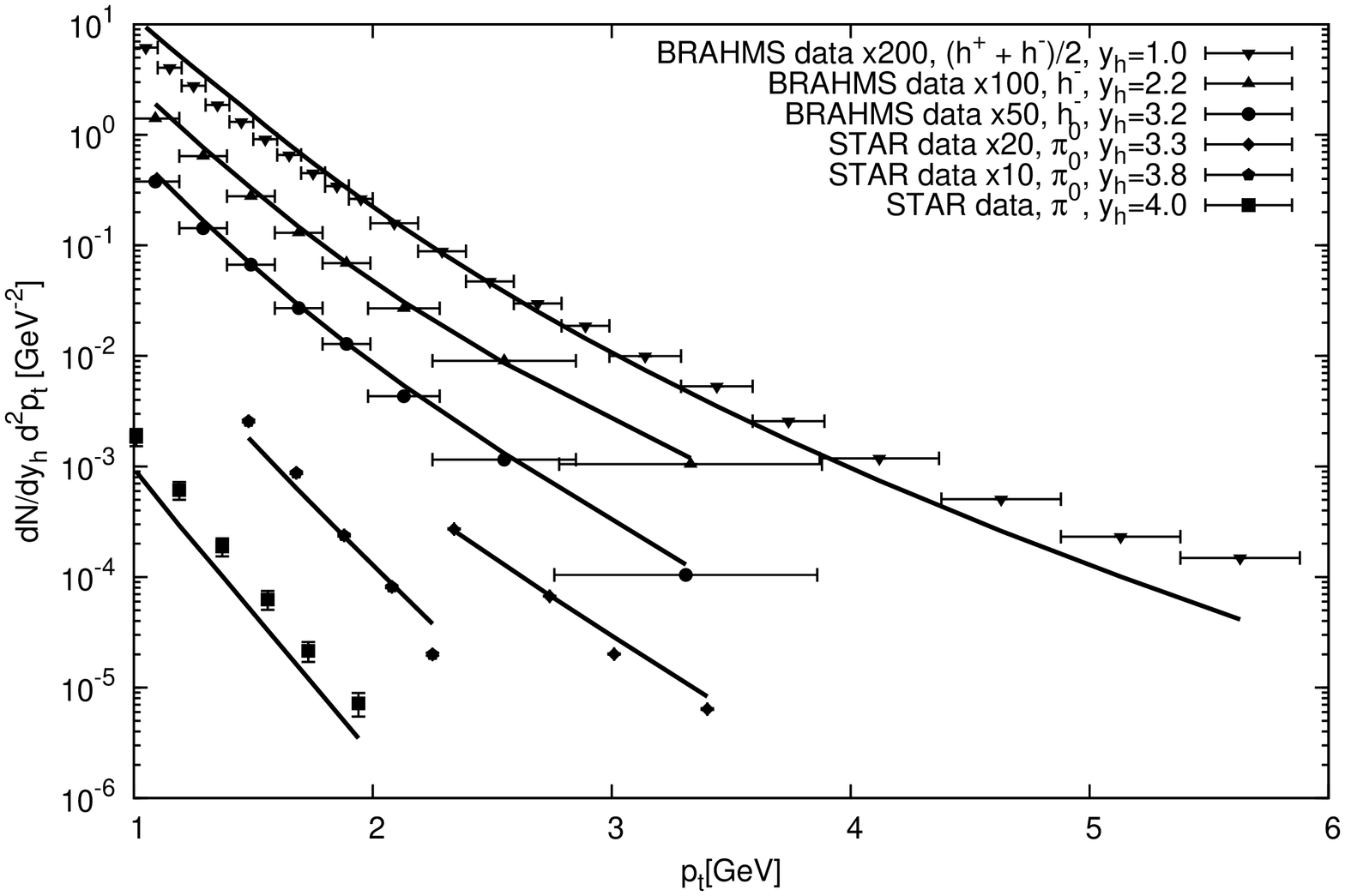}
\label{fig:agbs-pp-yh1}
}
\subfigure[$p+p$ collisions at RHIC for $y_h \geq 2.2$]{
\includegraphics[scale=0.33,angle=-90]{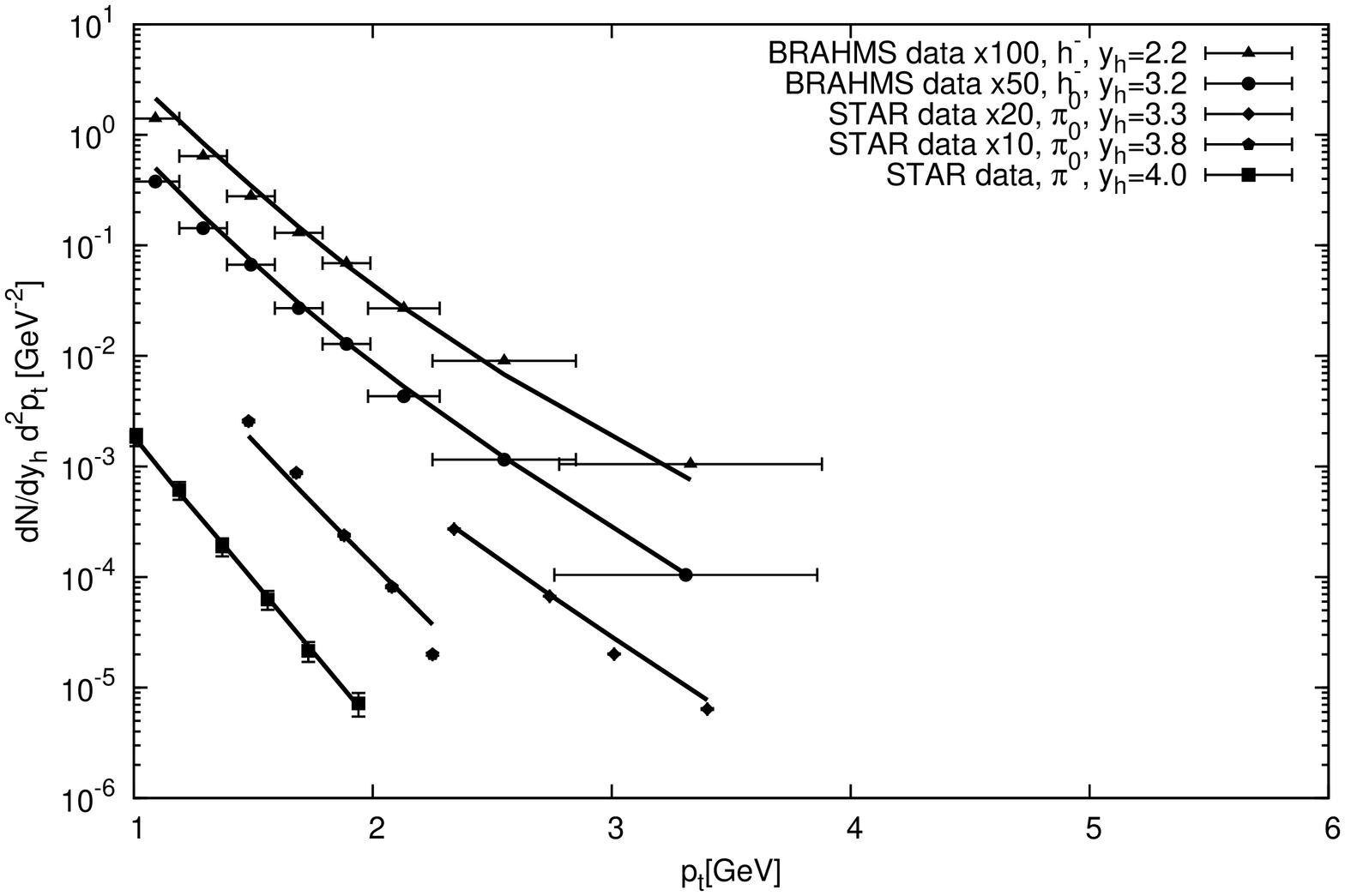}
\label{fig:agbs-pp-yh2}
}
\label{fig:agbs_rhic_pp}
\caption[]{Results for the RHIC charged hadron and $\pi^0$ yield for $p+p$ collisions from the simultaneous fit of AGBS to RHIC and HERA data.}
\end{figure}

The results for the simultaneous fit considering $p+p$ collisions at RHIC are presented in the Figs.\ \ref{fig:agbs-pp-yh1} and \ref{fig:agbs-pp-yh2}, where we see an even worse description in the fit of the mid-rapidity data in comparison with the $d+Au$ collisions. This is because the gold nucleus target has a greater partonic content than the proton at the same rapidity, being better described with the CGC formulation used here. The same does not happen with the proton target and corrections to Eq. (\ref{eq:had_yield}) should be important to describe this data at mid--rapidities. All in all, considering this fact and also remembering that AGBS is a model for the small-$x$ region of the QCD, the fit to the forward rapidities is the best choice for future use of the AGBS model in the single inclusive hadron production.

\subsection{LHC predictions}

\begin{figure}[t!]
\centering
\subfigure[$p+p$ collisions at LHC]{
\includegraphics[scale=0.33,angle=-90]{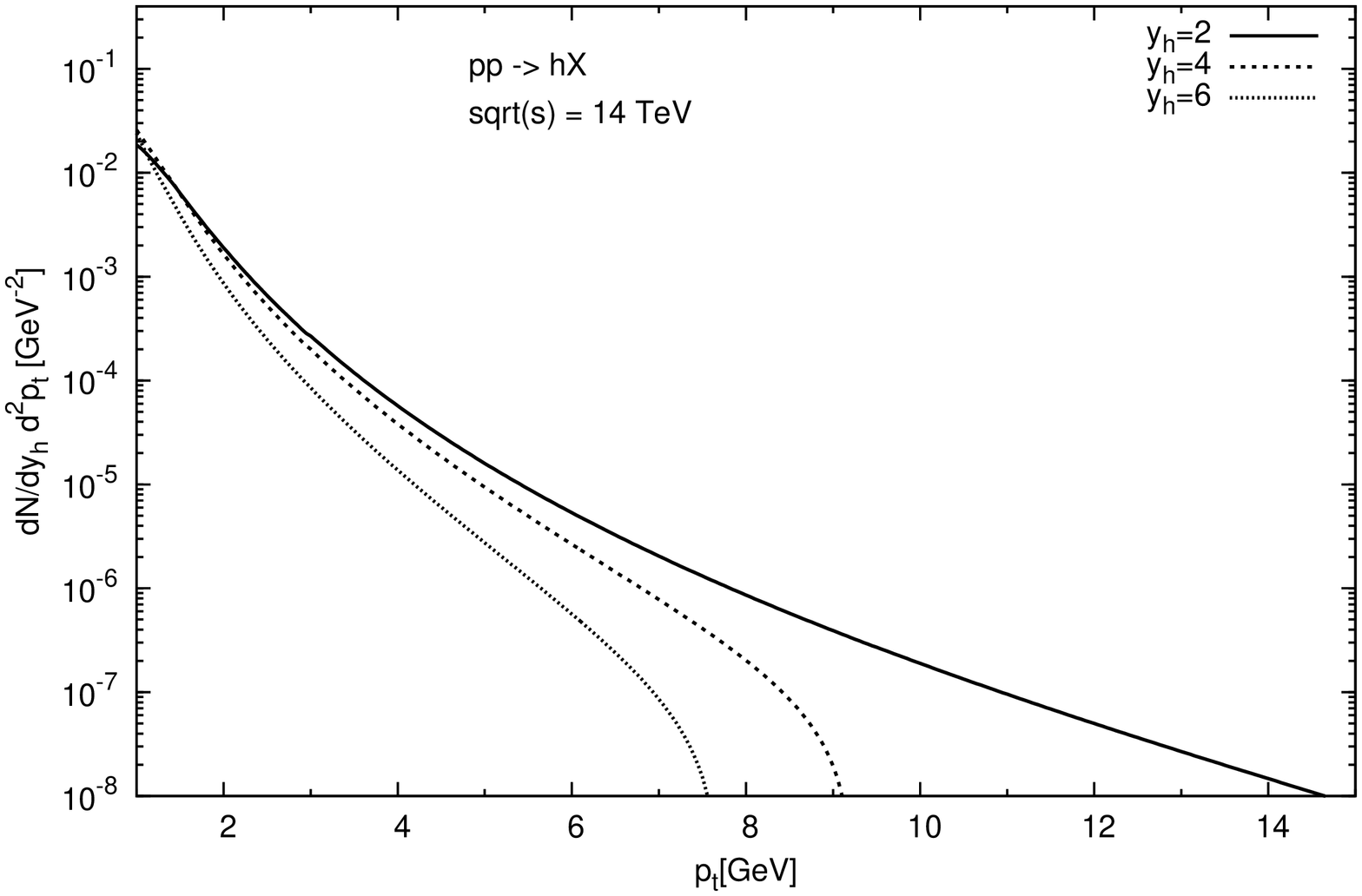}
\label{fig:agbs-lhc-pp}
}
\subfigure[$p+Pb$ collisions at LHC]{
\includegraphics[scale=0.33,angle=-90]{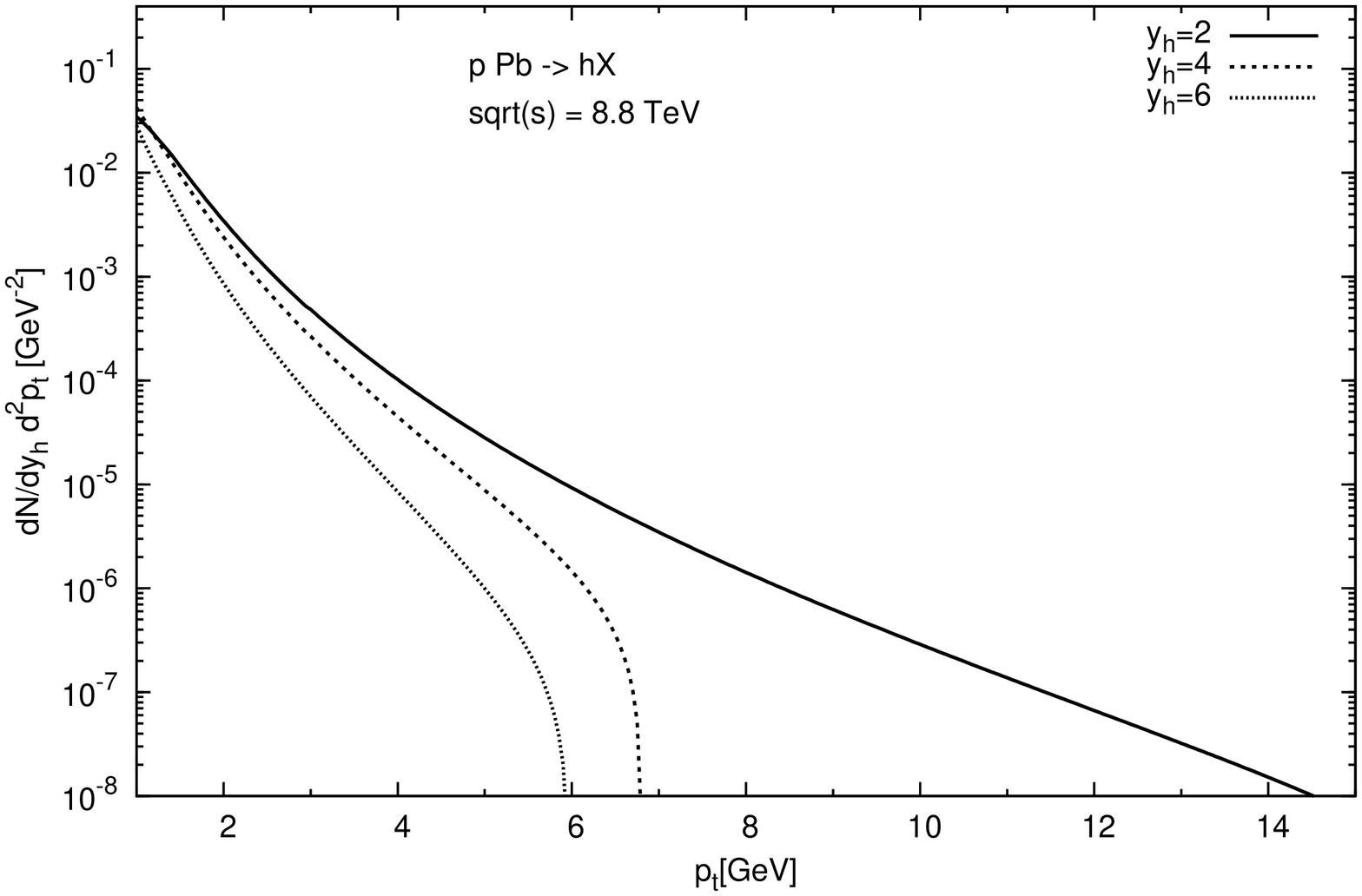}
\label{fig:agbs-lhc-pPb}
}
\label{fig:agbs_lhc}
\caption[]{Predictions of the AGBS model to the LHC charged hadron yield for $p+p$ and $p+Pb$ collisions. Parameters from the simultaneous fit of AGBS to forward rapidity RHIC and HERA data (first line of table \ref{tab:res}). We have used $A_\textrm{eff} = 20$ for lead target.}
\end{figure}

In this section we present predictions of the AGBS model for $p+p$ and $p+Pb$ collisions at LHC energies of 14 and 8.8 TeV, respectively. We use the parameters obtained from the simultaneous fit in the case of forward rapidity region of RHIC data (first line of the table \ref{tab:res}) as explained above and our results are shown in the Figs.\ \ref{fig:agbs-lhc-pp} and \ref{fig:agbs-lhc-pPb}. 

\begin{figure}[t!]
\centering
\includegraphics[scale=0.33,angle=-90]{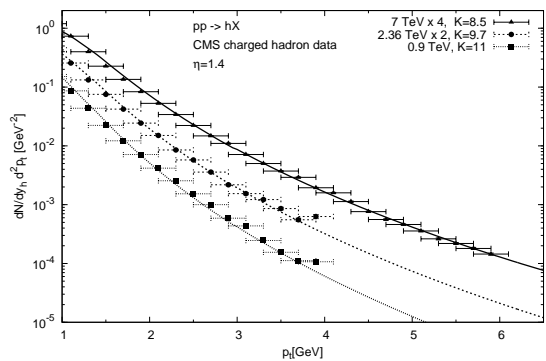}
\caption{Predictions of the AGBS model to the LHC CMS charged hadron yield for $p+p$ collisions at $\sqrt{s} = 0.9,\,2.36\,\mbox{and}\,7$ TeV. The experimental points are from CMS for $|\eta|<2.4$ \cite{cms-1,cms-2}. Parameters extracted from the simultaneous fit of AGBS to forward rapidity RHIC and HERA data (first line of table \ref{tab:res}).}\label{fig:cms-pp}
\end{figure}

Fig.\ \ref{fig:cms-pp} shows the description of the recent LHC CMS \cite{cms-1,cms-2} $p_t$ distribution data for proton--proton collisions using our model. 
We used the following expression to change the variable from rapidity to the measured pseudorapidity:
\begin{equation}
y(\eta,p_t,m) = \frac{1}{2}\ln\left[ \frac{\sqrt{m^2 + p_t^2 \cosh^2{\eta}} + p_t \sinh{\eta} }{\sqrt{m^2 + p_t^2 \cosh^2{\eta}} - p_t \sinh{\eta}} \right],
\end{equation}
where $m = 0.139$ GeV is the pion mass. We observed that for light hadrons ($m$ up to $0.4$ GeV), as in the case of the RHIC data, using $\eta \approx y_h$ does not affect considerably the results.

Surprisingly, a very good description was obtained, although the $K$ factors are large. This could be explained as an uncertainty of the AGBS model in the comparison with the pseudorapidity averaged data performed by the CMS collaboration. As the AGBS model is not supposed to describe the central rapidity region since it is a model for the low-$x$ part of the target (projectile), we used a averaged value $\eta = 1.4$ so that the description of the pseudorapidity averaged data over the region $|\eta|<2.4$ should imply some disagreements, which are in the $K$ factor. 

Standing for NLO corrections, the $K$ factors account for the $qq$ and $gq$ interactions, while Eq. (\ref{eq:had_yield}) considers only gluons in the target. 
Thus, as expected, the $K$ values decrease with the energy, because the target (projectile) gluon wavefunction is more important when compared with the quark density as the energy increases. In other words, Eq. (\ref{eq:had_yield}) should receive less quark corrections in the target for higher energies.

\section{Conclusions and discussion}\label{sec5}

We have shown in this work the compatibility of the AGBS model with both the single inclusive hadron yield at RHIC and the small-x DIS at HERA. Through Eq.\ (\ref{eq:agbs_nfs}) one could write analytically the AGBS model in the appropriate Fourier space used to describe the single inclusive hadron yield from the Color Glass Condensate. We have performed a new simultaneous fit of the AGBS model to the $F_2^p$ at HERA and hadron yield at RHIC, which results agreed with the one performed to HERA data alone, meaning that our expression correctly describes the AGBS in the standard Fourier space of the CGC formalism applied to hadron production. 

The model describes well the forward rapidity region of the RHIC data for $d+Au$ and $p+p$ collisions, but fails at mid-rapidities. Thus, being a model for the small-$x$ dipole amplitude, the parameters from the fit for the forward rapidities should be used to further employ of the AGBS model in inclusive hadron production.
The same formulation was used to make predictions to the LHC $p+p$ and $p+Pb$ collisions at 14 and 8.8 TeV, respectively. As a matter of data comparison, the model was also used to describe the CMS data on the hadron yield for $p+p$ collisions \cite{cms-1,cms-2} at different energies. Fig.\ \ref{fig:cms-pp} shows a pretty good description of the data, despite the high values of the $K$ factors.  As explained, this could be viewed as uncertainties of the model for the forward amplitude in the central region of pseudorapidity since the CMS data sums the entire region $|\eta|< 2.4$, and also in the target quark corrections not included in the CGC formulation for the hadron production. 

Although a good description of the nuclear data on inclusive hadron production was obtained, a study of the impact parameter dependence of the AGBS dipole amplitude is worth to investigate how sensitive in the transverse plane is the model. A better understanding of the parameters now averaged over the impact parameter would improve the model and its applicability to the high multiplicity events at LHC \cite{tv}. 

Furthermore, in order to get rid of the hadron fragmentation effects in our calculations, it is interesting to investigate prompt photon production in hadron collisions. As they interact only electromagnetically with the medium, the direct (prompt) photon arises as a good tool to investigate initial state effects and saturation effects in hadronic collisions. This observable is well studied in the color dipole formalism \cite{hq+photon} and could be also used to investigate the heavy quark contribution to the dipole scattering amplitudes through the photon plus heavy quark production, as recently done in \cite{bm-2010} with both dipole models \cite{dhj2,buw}.

At last, a next-to-leading order (NLO) phenomenological modeling of the traveling wave QCD method is needed to mark the region of applicability of the AGBS model. The theory of the method at higher orders has already been done, in the spirit of regularization group scheme \cite{bp,ps,enberg}. However there are no applications in phenomenology yet, being an interesting field for future studies.

\section*{Acknowledgements}
This work is supported by CNPq (Brazil). E.G.O.\ is supported by CNPq under contract 201854/2009-0.

\bibliographystyle{unsrt}
 
\end{document}